# Selectivity in single-molecule reactions by tip-induced redox chemistry

Florian Albrecht[1, *, †], Shadi Fatayer[1, 2, †], Iago Pozo[3, #], Ivano Tavernelli[1], Jascha Repp[4], Diego Peña[3, *], Leo Gross[1, *]

[1] IBM Research Europe – Zurich, 8803 Rüschlikon, Switzerland

[2] Applied Physics Program, Physical Science and Engineering Division, King Abdullah University of Science and Technology (KAUST), Thuwal 23955-6900, Kingdom of Saudi Arabia

[3] Centro Singular de Investigación en Química Biolóxica e Materiais Moleculares (CiQUS) and Departamento de Química Orgánica, Universidade de Santiago de Compostela, 15782- Santiago de Compostela, Spain

[4] Institute of Experimental and Applied Physics, University of Regensburg, 93053 Regensburg, Germany

[#] Present address: Chemistry Research Laboratory (CRL), Department of Chemistry, University of Oxford, Oxford, OX1 3TA, UK

[†] These authors contributed equally to the work

* Corresponding authors. Email: FAL@zurich.ibm.com, diego.pena@usc.es, LGR@zurich.ibm.com

**Abstract:** Controlling selectivity of reactions is a quest in chemistry. Here, we demonstrate reversible and selective bond formation and dissociation promoted by tip-induced reduction-oxidation reactions on a surface. Molecular rearrangements leading to different constitutional isomers are selected by the polarity and magnitude of applied voltage pulses from the tip of a combined scanning tunneling and atomic force microscope. Characterization of voltage dependence of the reactions and determination of reaction rates demonstrate selectivity in constitutional isomerization reactions and provide insight into the underlying mechanisms. With support of density functional theory calculations, we find that the energy landscape of the isomers in different charge states is important to rationalize the selectivity. Tip-induced selective single-molecule reactions increase our understanding of redox chemistry and could lead to novel molecular machines.

**One-Sentence Summary:** Different bonds within a single molecule can be formed selectively by atom manipulation through the voltages applied.

Controlling selectivity in reactions is a central goal in chemistry. In solution, such control can be achieved in steering the valence electrons by the pH-value or the electrochemical potential, for example. By these means, however, the reaction conditions are altered in such complexity that the basic mechanisms governing selectivity often remain elusive. The exploration of how external electrostatic fields and charge-state manipulation affect chemical bonding is still in its infancy [1].

Investigating surface chemistry by means of STM and AFM offers the possibility to study basic chemical mechanisms under atomically well-defined conditions. The reaction itself can be even directly triggered with the tip of an STM at will [2,3]. Spurred by advances in molecular characterization by STM [4] and AFM [5], novel tip-induced reactions and reaction mechanisms were discovered [6-11]. Typically, in tip-induced chemistry, precursors are designed on which specific bonds can be broken, dissociating designated masking groups [3]. The demasking can in turn enable or cause other reactions such as intermolecular bond formation [3,11,12], intramolecular bond formation [7] or skeletal rearrangements [13].

Control over configurational switching [14], bond formation and dissociation [15] and hydrogen tautomerization reactions [16-18] could be achieved by means of charge attachment and charge-state manipulation. By means of the electric field, control of configurational isomers [19] and control of the yield of a Diels-Alder reaction [20] were demonstrated. Furthermore, selectivity between molecular translation and desorption, controlled by inelastically tunnelling electrons [21] and selective bond dissociation resulting from adsorbate-substrate bond alignment [22] were achieved. Even tip-controlled artificial molecular machines [23] have been demonstrated. For example, a molecule translated by molecular motors driven by inducing alternating conformational and configurational isomerization reactions [24].

Here, we showcase the potential of tip-induced electrochemistry obtaining chemical selectivity in single-molecule reactions, that is, we show that multiple constitutional isomerization reactions can be controlled and selected by voltage pulses from the tip. Selected by the voltage pulse, we formed different transannular covalent bonds.

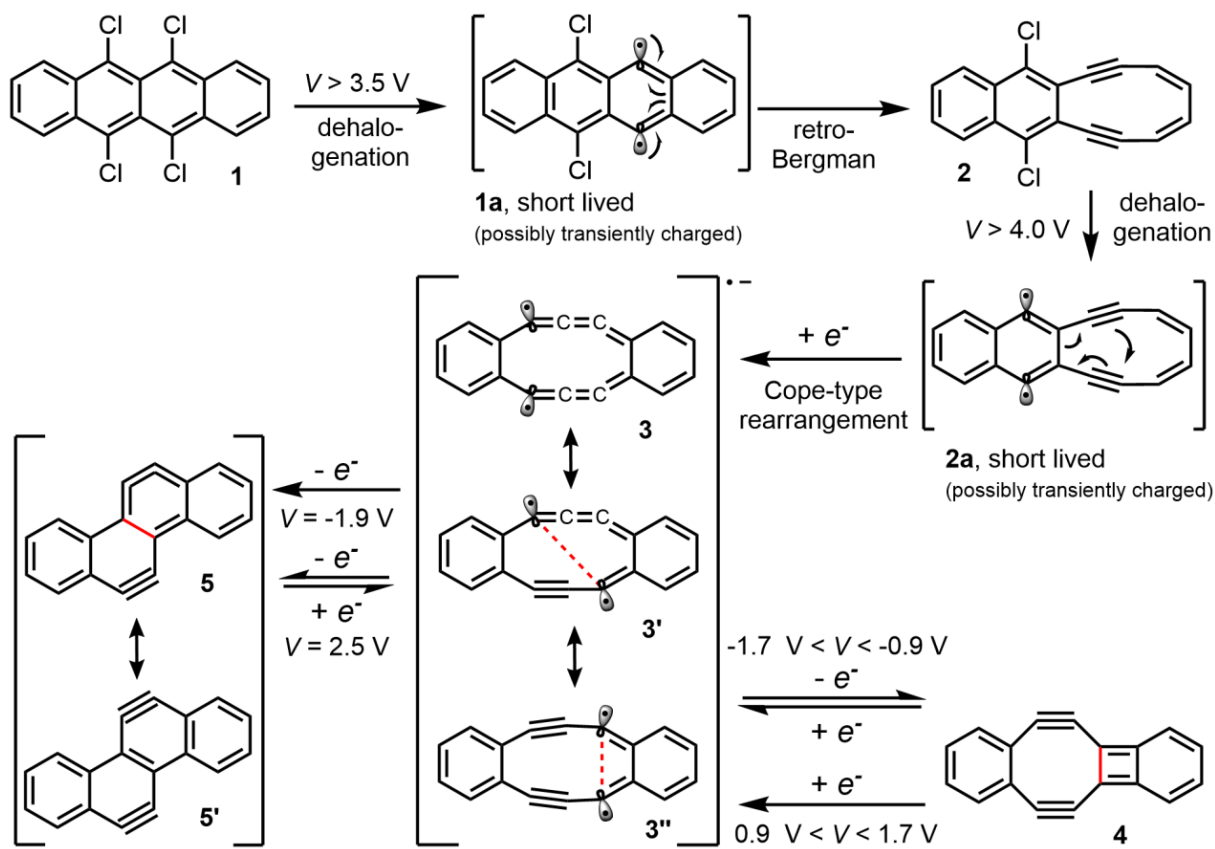

**Fig. 1: Reaction scheme.** Tip-induced reactions promoted by voltage pulses are indicated with the respective sample voltages *V*. For **3** and **5** different resonance structures are displayed. The intermediates **1a** and **2a** might be transiently charged. We also observed other partly dehalogenated intermediates than **2** (see fig. S3), indicating several pathways for the initial formation of **3**, **4** and **5**.

As molecular precursor we synthesized 5,6,11,12-tetrachlorotetracene (**1**, $C_{18}H_8Cl_4$, see Fig. 1 and Supplemental Material, Scheme S1 and S2, for details on the synthesis). We deposited **1** by thermal sublimation on a Cu(111) substrate partly covered with NaCl islands of one to three monolayers (ML) thickness, at a sample temperature of $T \approx 10$ K. Experiments were performed at $T = 5$ K, on molecules on 2ML NaCl on Cu(111) unless noted otherwise. All images reported were obtained with CO-functionalized tips [5]. All AFM images are recorded in constant-height mode at a sample voltage of $V = 0$ V.

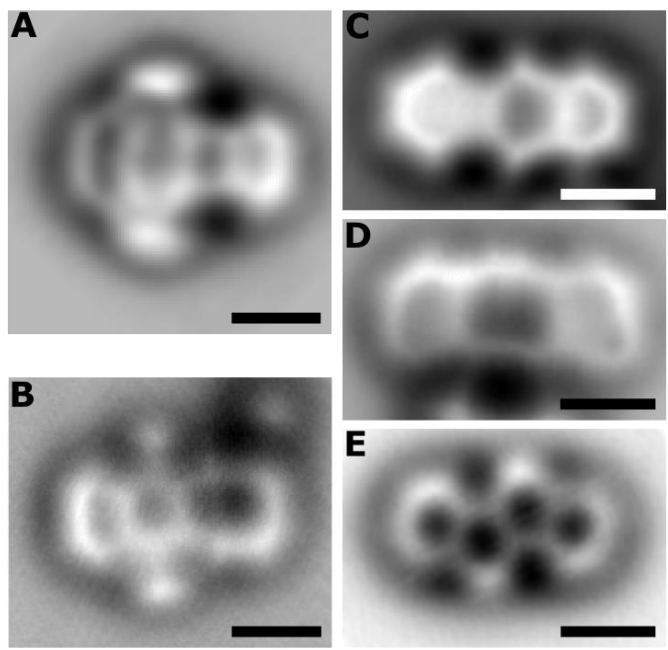

**Fig. 2**: **AFM images of precursor, intermediate and products.** Constant-height, CO-tip AFM images of precursor **1** (**A**), intermediate **2** (**B**) and products **4** (**C**), **3** (**D**) and **5** (**E**), on NaCl. (A, B, D) on 2ML NaCl, (C, E) on 1ML NaCl. Scale bars correspond to 0.5 nm. For images at different tip-height offsets and imaging parameters, see fig. S1, S2, S4, S5, S6.

Figure 2A shows an AFM image of **1**. The differences in brightness of the Cl atoms result from steric hindrance between neighboring Cl atoms, causing different adsorption heights see fig. S1. By means of voltage pulses from the probe tip located above the molecule, we dissociated Cl atoms. We observed a threshold voltage of about $V = +3.5$ V, with tunneling currents on the order of $I = 1$ pA, for the dissociation of the first two Cl atoms from **1**. Often, the molecule also moved on the surface by a few nm when a voltage pulse was applied.

Figure 2B and fig. S2 show the partly dechlorinated intermediate **2** ($C_{18}H_8Cl_2$), with two Cl atoms of **1** dissociated (see other partly dechlorinated intermediates observed in fig. S3 and Table S1). The AFM image of **2** reveals a ten-membered ring on the dechlorinated side, exhibiting characteristic sharp and bright features above the triple bonds [7]. This suggests that a retro-Bergman-cyclization reaction [7] had taken place with intermediate **1a** as transition structure. By voltage pulses of +4 V to +4.5 V, we dissociated the remaining chlorine atoms in **2**, creating different structures with the chemical formula $C_{18}H_8$. We observed constitutional isomers **3**, **4** and **5**, shown in Fig. 2C-E (see also fig. S4, S5 and S6) and in rare occasions other isomers (see fig. S7).

These molecules are highly strained, presumably very reactive and none of them had been reported previously. Because of the inert NaCl surface and the low temperature they are stable under the conditions of our experiment for $|V| < 0.7$ V. The isomers differ in their central part, where they either exhibit a ten-membered ring in **3**, a four-membered ring fused to an eight-membered ring in **4**, or two fused six-membered rings, yielding a chrysene-like carbon backbone in **5**. In most cases (62%) we found structure **3** directly after the dissociation of all chlorine atoms (see Table S2).

Figure 1 shows a plausible synthetic route for the formation of these isomeric strained hydrocarbons. Probably, the cleavage of two Cl atoms from intermediate **2**, at first, generates sigma diradical **2a**, which could evolve through a Cope-type rearrangement to obtain structure **3**. Possibly this reaction proceeds in a transiently charged state (see SM text: Details on dehalogenations). This would be in accordance with the mechanism previously proposed for enediyne cyclizations, which is facilitated by the formation of radical-anionic species [25]. While **3** is a plausible resonance structure with two cumulene moieties, alternative structures combining enyne and cumulene groups (**3'**) or two enynes (**3''**) within the central ten-membered ring can also be considered. The transannular C-C bond formation between the radicals in structure **3''** would lead to the formation of diyne **4**, while the C-C bond formation between radicals in **3'** would afford the chrysene-based bisaryne **5**.

First, we characterize on surface the products **3**, **4** and **5**. The STM measured maps of electronic resonances are shown in Fig. 3 and fig. S8, accompanied by DFT calculations. We find that **4** and **5** are charge neutral, with the shape of the imaged frontier orbital densities, of the highest occupied and lowest unoccupied molecular orbital (HOMO and LUMO) in agreement to the DFT calculations, see Fig. 3A-K and fig. S9. For **4** we did not observe the positive ion resonance (HOMO) for voltages up to $V = -2$ V, see fig. S10. For **3** the experiment shows that the molecule is in its anionic charge state. This is consistently indicated the scattering of interface-state electrons (fig. S11), Kelvin probe force spectroscopy (Table S3) and STM maps of the electronic resonances (Fig. 3B, C) and their comparison with theory (figs. S12 and S13). The identification of structure **3** as a radical anion agrees with the expected tendency of sigma radicals to be reduced, in contrast to closed-shell compounds **4** and **5**.

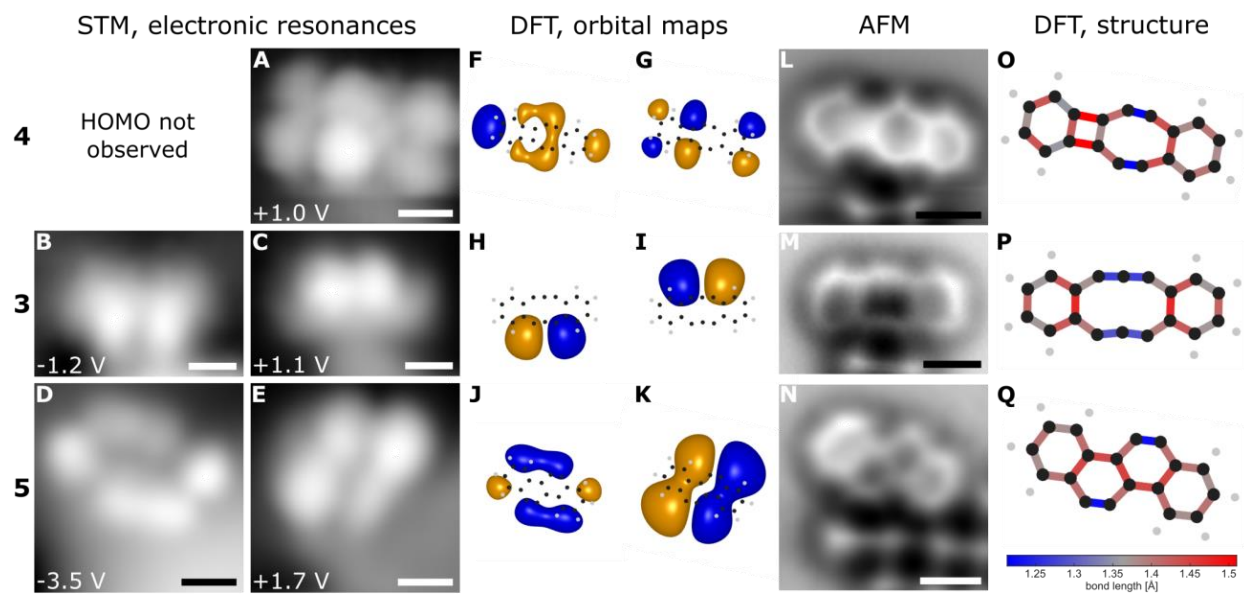

**Fig. 3**. **Characterization of products by STM, DFT and AFM**. (**A-E**) STM images at the first electronic resonances at negative and positive sample voltage. (**F-K**) Simulated orbital maps of the frontier orbitals, considering the finite resolution of the tip (see fig. S9). (**L-N**) AFM images. All experimental images were recorded on 2ML NaCl, with the molecules next to 3ML NaCl islands. (**O-Q**) DFT calculated structures on NaCl with the bond length indicated by color. The DFT calculations of **3** are shown in the anionic charge state, that for **4** and **5** in the neutral charge state. Scale bars correspond to 0.5 nm.

Moreover, our measurements reveal that the anion of **3** undergoes a Jahn-Teller distortion that is also found by B3LYP-based DFT (see Fig. 3P) and explains the symmetry breaking with respect to the long molecular axis that is observed for both electronic resonances (Fig. 3B, C). Structurally, the Jahn-Teller distortion of the anion **3$^{-1}$** features an inward bend on one long side of the ten-membered ring (see Fig. 3P) and can be observed in high-resolution AFM images (see Fig. 3M and fig. S4). The first electronic resonance at negative bias, shows two lobes of increased orbital density on the side of the inward bend (see Fig. 3B, H and fig. S12). Whereas the first electronic resonance at positive bias exhibits increased density on the side opposite to the inward bend (see Fig. 3C, I and fig. S12 and S13) in excellent agreement of experiment and theory.

Next, we study tip-induced reactions between **3**, **4** and **5**. When we applied pulses of relatively large bias, $V > +2.5$ V with currents $I$ on the order of 10 pA, we could transform molecules between all these three structures, however, with limited control of the outcome. A rearrangement after a bias pulse of $V = +2.5$ V resulted mostly in structure **5** (about 50% of the attempts) and less often in structures **4** or **3** (25% each). Structure **5** was stable for $|V| < 2$ V. However, voltage pulses of $|V| < 2$ V, when applied to **3** and **4**, resulted in different reactions depending on the applied voltage.

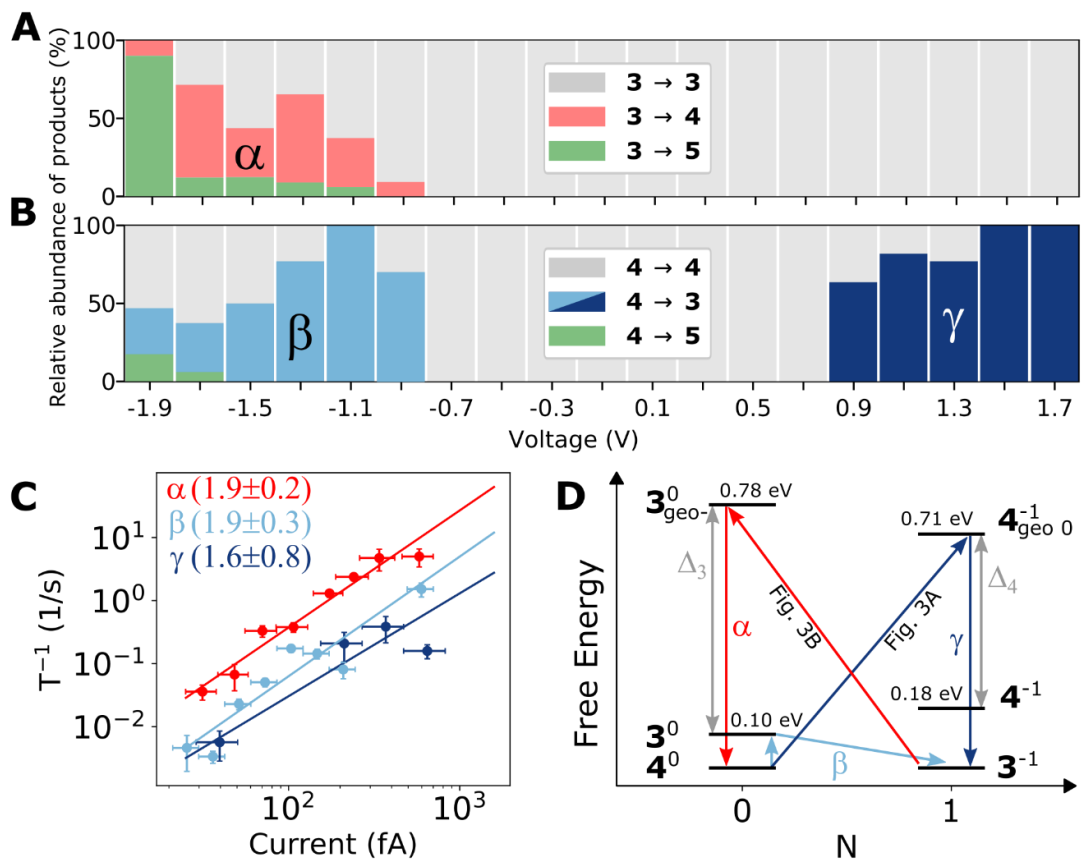

**Fig. 4**. **Tip-induced transitions between 3 and 4.** Histograms (**A**, **B**) show the outcome in dependence of $V$ applied for $t = 10$ s, at constant current $I = 0.5$ pA above molecule **3** (A) and **4** (B) on 2ML NaCl. (**C**) Reaction rate as a function of tunneling current. Transitions α and β were probed at $V = -1.3$ V, transition γ was probed at $V = 1.1$ V. The extracted slopes are for α = 1.86 ± 0.18, β = 1.90 ± 0.26, γ = 1.61 ± 0.76. (**D**) Energy level diagram for $V \approx 0$ V, associated with the observed charging and structural transitions between **3** and **4**, assuming equal free energies of **4**$^0$ and **3**$^{-1}$.

Histograms showing the outcome of voltage pulses with currents of $I = 0.5$ pA applied above **3** and **4** as initial structures are shown in Fig. 4A, B, respectively (see also fig. S14 and fig. S15). Our findings are summarized as: (*i*) At $|V| < 0.7$ V, **3** and **4** are stable. (*ii*) At -1.7 V $< V <$ -0.9 V, bidirectional switching occurs between **3** and **4**. That is, **3** can be converted into **4** (transition labelled α) and *vice versa* (transition β), and only with a small probability (< 10 %) structure **5** is formed. (*iii*) At +0.9 V $< V <$ +1.7 V, unidirectional switching from **4** to **3** occurs. That is, **4** is converted into **3** (transition γ), but **3** is stable at these voltages. (*iv*) At $V \approx$ -1.9 V structure **3** is transformed into **5** with a high yield and structure **4** either into **3** or **5**.

Switching between the structures was possible on 1, 2 and 3 ML NaCl islands with similar threshold values. The data shown in Fig. 4 was measured on 2 ML NaCl. The observed voltage dependence allows us to select outcomes of tip-induced rearrangements. We can select which transannular C-C bond is formed within the ten-membered ring of structure **3**. We dominantly generated structure **4** with pulses in the range of -1.1 V to -1.7 V, but structure **5** with $V = -1.9$ V, demonstrating selectivity in single-molecule constitutional isomerization reactions.

For the transitions between **3** and **4** we investigated the respective reaction rates, see Fig. 4C. These were measured at $V = -1.3$ V for transitions α (from **3** to **4** at negative *V*) and β (from **4** to **3** at negative *V*), and at $V = +1.1$ V for γ (from **4** to **3** at positive *V*), as a function of current by using different tip heights. The slopes of the linear fits in the double logarithmic plot are 1.86 ± 0.18 for α, 1.90 ± 0.26 for β, and 1.61 ± 0.76 for γ. This indicates that transitions α and β are two-electron processes. For γ the error is too large to differentiate between a one- and a two-electron process.

Figure 4D visualizes the transitions between **3** and **4**. The transition α and γ coincide with the onset of ionic resonances of the initial structures, as probed by STM (Fig. 3B, A, respectively), suggesting that these transitions involve (de)charging the initial structure. Note that at electronic resonance, the charge state is transiently changed by charge transfer between tip and molecule, after a typical lifetime on the order of few ps on 2ML NaCl the molecular charge ground state is restored by charge transfer between molecule and metal substrate [4]. The structural relaxation that follows a charge transition can oscillate for several 10 ps [26].

For Fig. 4D we calculated on a NaCl surface the ground state energies of **3** and **4** in different charge states and the related relaxation energies Δ (see SM) [27]. The relatively large energy of the intermediate $3^0_{geo-}$ for the transition from $3^{-1}$ is rationalized by the Jahn-Teller distortion of $3^{-1}$. The calculated energies for the charge transitions are in good agreement with the resonances measured by STM when assuming similar energies for $4^0$ and $3^{-1}$. A partial voltage drop across NaCl of about 20% can be considered for this junction geometry [27].

The observed two-electron process for α indicates that in addition to the charge transition a second charge carrier is needed to provide additional energy in an inelastic electron tunneling (IET) process [28]. In contrast, transition β involves no charging of the initial structure $4^0$, but probably only by IET processes a transition to $3^0$ occurs, which is subsequently charged from the substrate to $3^{-1}$, the charge ground-state of structure **3**. In transition γ, the LUMO of **4**, shown in Fig. 3G, is transiently occupied. This orbital exhibits a nodal plane along the long axis of the molecule, thus is an antibonding state with respect to the central bond of **4**, facilitating its rupture [29] and thus the transition to **3**. Note that at small currents of $I \approx 0.5$ pA the reaction rates are on the order of minutes, and thus orbital density images can be obtained, see Fig. 3A, B, C. Our results indicate that these reactions are triggered by electron attachment, rather than by the electric field alone. That the effect of the latter is small can be rationalized by the reaction coordinates, i.e., the molecular plane and the movement of atoms, being parallel to the substrate, whereas the electric field applied by the tip is orthogonal with respect to the substrate and the reaction coordinates.

On-surface calculations (see fig. S16) show that in both the neutral and negative charge state, **5** has a lower energy than **3** and **4**, explaining the observed dominant switching to **5** at larger biases and its stability for $|V| < 2.0$ V. Gas-phase calculations (see fig. S16) indicate that the reaction barriers (both in neutral and negative charge states) to **5** are higher than the ones between **3** and **4**, explaining the increased voltages needed to switch to **5**, compared to switching between **3** and **4**. The high barrier between **4** and **5** in the neutral and negative charge states (see fig. S16) suggests that the observed transformation from **4** to **5** at $V = -1.9$ V (see Fig. 4B) proceeds via structure **3** as intermediate.

The switching at higher bias $|V| > 2.0$ V and increased currents on the order of 10 pA is less controlled and more challenging to understand and its detailed description is beyond the scope of this work. Because of the large bias, several successive and branching charge transitions and structural rearrangements must be considered. Also, current induced catalytic reduction of the barrier might play a role [30]. In addition, the switching between all three structures at $V = 2.5$ V could include higher excited states, e.g., occupation of the LUMO+1, which for **5** is accessed at $V = 2.5$ V, (see fig. S17). It might proceed via dianionic charge states, which show similar ground state energies for all three structures (see fig. S16), and the transient occupation of the dianion is observed for **3** by resonant tunneling in Fig. 3C at $V = 1.1$ V.

Our experiment shows that for a molecule on a surface, several chemical transformations between multiple constitutional isomers can be controlled by tip-induced redox chemistry. With different voltages and polarities, we selectively activated one, two or all three transitions between three different isomers (**3**, **4**, **5**). We demonstrated directed and reversible switching between two non-ground state isomers (**3**, **4**) and selectively formed transannular covalent bonds deliberately transforming **3** to either **4** or **5**. We learned that the selectivity of the reactions is facilitated by changes of the energy landscape as a function of the transient charge state, accessed by the bias applied. The charge ground states can in general be adjusted via the substrate's work function [31].

The insights obtained in redox reactions studied by tip-induced electrochemistry, will be useful to better understand redox reactions that are important in organic synthesis [32] and nature [33]. For future artificial molecular machines [23], controlled, reversible and selected switching between more than two different constitutional isomers, as demonstrated in our work, could enable novel functionalities. In addition, increased workload and operation at elevated temperatures could be facilitated by the relatively high energy barriers, on the order of 1 eV, involved in constitutional isomerization reactions.

**Acknowledgments:** We thank Rolf Allenspach, Shantanu Mishra, Enrique Guitián and Zhiyong Zhu for discussions. This work was supported by the ERC Synergy Grant MolDAM (no. 951519), the ERC Consolidator Grant AMSEL (no. 682144) and the European FET-OPEN project SPRING (no. 863098), the Spanish Agencia Estatal de Investigación (PID2019-107338RB-C62, PID2019-110037GB-I00 and PCI2019-111933-2), Xunta de Galicia (Centro de Investigación de Galicia accreditation 2019–2022, ED431G 2019/03) and the European Regional Development Fund-ERDF. For some of the calculations, this research used the resources of the Supercomputing Laboratory at King Abdullah University of Science & Technology (KAUST) in Thuwal, Saudi Arabia.

**Funding:**

    ERC Synergy Grant MolDAM (no. 951519)

    ERC Consolidator Grant AMSEL (no. 682144)

    European FET-OPEN project SPRING (no. 863098)

    Spanish Agencia Estatal de Investigación (PID2019-107338RB-C62)

    Spanish Agencia Estatal de Investigación (PID2019-110037GB-I00)

    Spanish Agencia Estatal de Investigación (PCI2019-111933-2)

    Xunta de Galicia (Centro de Investigación de Galicia accreditation 2019–2022, ED431G 2019/03)

**Author contributions:**

    On-surface experiments: FA, SF, LG

    Synthesis of precursor molecules: IP, DP

    DFT calculations: SF, IT, IP, FA, LG

    Writing: LG, FA, SF, IP, IT, JR, DP

**Competing interests:** Authors declare that they have no competing interests.

**Data and materials availability:** All data are available in the main text or the supplementary materials.

**Supplementary Materials**

Materials and Methods

Supplementary Text

Figs. S1 to S17

Table S1 to S3

References (*34–49*)

# Supplementary Materials for

## Selectivity in single-molecule reactions by tip-induced redox chemistry

Florian Albrecht[1, *, †], Shadi Fatayer[1, 2, †], Iago Pozo[3, #], Ivano Tavernelli[1], Jascha Repp[4], Diego Peña[3, *], Leo Gross[1, *]

[1] IBM Research Europe – Zurich, 8803 Rüschlikon, Switzerland

[2] Applied Physics Program, Physical Science and Engineering Division, King Abdullah University of Science and Technology (KAUST), Thuwal 23955-6900, Kingdom of Saudi Arabia

[3] Centro Singular de Investigación en Química Biolóxica e Materiais Moleculares (CiQUS) and Departamento de Química Orgánica, Universidade de Santiago de Compostela, 15782-Santiago de Compostela, Spain

[4] Institute of Experimental and Applied Physics, University of Regensburg, 93053 Regensburg, Germany

[#] Present address: Chemistry Research Laboratory (CRL), Department of Chemistry, University of Oxford, Oxford, OX1 3TA, UK

[†] These authors contributed equally to the work

* Corresponding authors. Email: FAL@zurich.ibm.com, diego.pena@usc.es, LGR@zurich.ibm.com

**This PDF file includes:**

Supplementary Text
- Solution synthesis of 5,6,11,12-tetrachlorotetracene (**1**)
- Experimental details for on-surface experiments
- Details on DFT calculations

Figs. S1 to S17

Tables S1 to S3

**Solution synthesis of 5,6,11,12-tetrachlorotetracene (1)**

5,6,11,12-Tetrachlorotetracene (**1**) was obtained in two steps (Schemes S1 and S2) from commercially available 6,11-dihydroxy-5,12-tetracenedione (**6**) following a previously reported procedure [34].

To a mixture of PCl$_5$ (50 mg) and compound **6** (Scheme S1, 200 mg) in a flame-dried round-bottom flask, POCl$_3$ (2.6 mL) was slowly added using a dropping funnel. The red mixture was refluxed (approximately 110 °C) for 5 h until the red color had dissipated. The reaction mixture was cooled to room temperature and filtered through a fritted glass filter. The precipitate was washed sequentially with glacial acetic acid and hexanes. The residue was collected and dried under vacuum to give **7** as a white powder (240 mg, 80%). $^1$H NMR (300 MHz, CDCl$_3$): δ 8.69 (dd, $J$ = 3.3, 6.6 Hz, 2H), 8.14 (dd, $J$ = 3.3, 6.3 Hz, 2H), 7.80 (dd, $J$ = 3.3, 6.6 Hz, 2H), 7.55 (dd, $J$ = 3.3, 6.3 Hz, 2H) ppm.

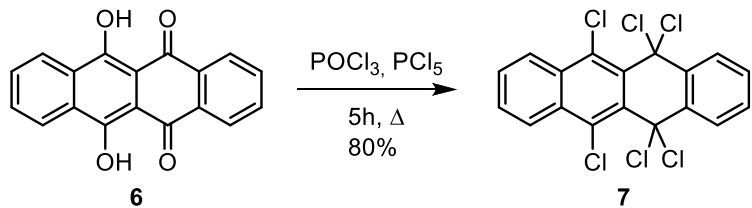

**Scheme S1.** Synthesis of 5,5,6,11,12,12-hexachloro-5,12-dihydrotetracene (**7**).

A solution of **7** (200 g, Scheme S2) and NaI (295 mg) in DMF (2.7 mL) was refluxed under Ar atmosphere for 30 min, cooled to 80 °C, and filtered warm through a fritted glass filter. The precipitate was washed sequentially with ethanol, water, and acetone. The residue was dried under vacuum to give **1** as a red solid (130 mg, 79%). $^1$H NMR (300 MHz, CDCl$_3$): δ 8.60 (dd, $J$ = 3.3, 6.9 Hz, 4H), 7.62 (dd, $J$ = 3.3, 6.9 Hz, 4H) ppm. Spectroscopic data matched with those reported in the literature.

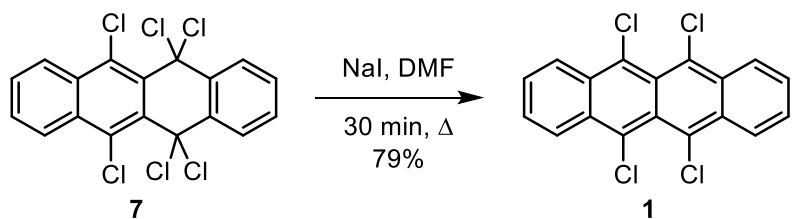

**Scheme S2.** Synthesis of 5,6,11,12-tetrachlorotetracene (**1**).

## Experimental details for on-surface experiments

On-surface experiments were carried out in a homebuilt, ultra-high vacuum, combined STM and AFM operated at a temperature of 5 K. The Cu(111) surface was cleaned by several sputtering and annealing cycles and the NaCl films were deposited at a sample temperature between 230 K and 330 K resulting in islands of one to three monolayer thickness. Precursors **1** were thermally deposited onto the cold ($T \approx 10$ K) surface using an oxidized piece of silicon wafer, by direct current heating. CO molecules for tip functionalization were dosed from the gas phase onto the cold sample. The scanner was equipped with a qPlus sensor [35] and operated in frequency-modulation mode [36] with a constant oscillation amplitude of $A = 0.5$ Å. The bias voltage $V$ was applied to the sample with respect to the tip. AFM images were recorded in the constant-height mode, at $V = 0$ V, with the tip offset $\Delta z$ applied with respect to the STM controlled setpoint above the bare NaCl/Cu(111) surface. Negative (positive) $\Delta z$ corresponds to a decrease (increase) in tip-sample distance with respect to the setpoint. All AFM and STM images reported were obtained with CO-functionalized tips [5].

To investigate the reaction rates of the chemical transitions as a function of current (see manuscript Fig. 4C), we recorded current traces without $z$-feedback observing $N = 440$ chemical transformations. We recorded 195 times a transformation α, 205 times β and 40 times γ. For γ we probed less transitions, because the experimental effort is significantly higher for probing this transition. Because at positive $V$ the reaction is unidirectional from **4** to **3** (transformation γ), we had to apply a reset pulse at negative $V$ to reset to **4** before probing another event of transformation γ. However, the reset pulse at negative voltage can induce both reactions α and β, requiring high-resolution AFM measurements to confirm a successful reset by reaction α to **4**. For one of the molecules, we observed switching without lateral displacement. In this case, we could record current traces $I(t)$ at negative bias voltage with numerous transitions α and β (see fig. S14).

## Details on dehalogenations

In contrast to the selected, directed and reversible reactions between structures **3**, **4** and **5**, their initial formation from **1** is more challenging to understand. This is because the initial formation is not reversible and required pulses of larger voltages allowing many possible charge and structure transitions and is therefore less controlled. In this section we discuss the reaction pathway from **1** towards **3**, **4** and **5**, based on 163 completed dehalogenations of **1**.

In addition to the intermediate **2**, several other partly dehalogenated intermediates were observed after applying voltage pulses of +3.5 V to +4.5 V above **1**, see fig. S3. Observing these different intermediates indicates that there exist several reaction pathways for the creation of **3**, **4** and **5** from **1** on the surface, in addition to the reaction pathway via intermediate **2** shown in Fig. 1.

During the experiments in total $N = 48$ partly dehalogenated intermediates were observed and we detected partly dehalogenated intermediates with three, two, and one Cl atoms still attached, with relative occurrences of 6%, 81% and 13%, respectively. The occurrences indicate that the Cl atoms are in most cases - but not exclusively - dissociated pairwise. This indicates smaller energy barriers for the removal of the second (fourth) Cl atom, compared to the first (third) Cl atom.

There are four different possible intermediates observed with two Cl atoms, see fig. S2 and fig. S3 A, B, C, that is, structures **2**, **1a**, **1b**, **1c**, respectively. The most frequently observed intermediate was **1c**. This could be a result of **1c** being the dominantly formed intermediate, however, it could also be the result of **1c** having the highest barrier for the reaction to proceed. That is, intermediates **1a**, **1b**, and **2** might have transformed with higher probabilities than **1c** directly into **3**, **4** or **5** without the need of a second voltage pulse. In about 60% of the cases **1** was transformed into **3**, **4** or **5** by a single voltage pulse of 3.5 to 4.5 V, without observing a partly dehalogenated intermediate.

For the intermediates **1a**, **1b**, **1c** and **2**, we calculated the energies in the gas phase in different charge states, see Table S1. We observe that **2** has the lowest energy in the neutral and negative charge state, and **1b** has the lowest energy in the doubly negative charge state. The most frequently observed intermediate with two Cl atoms **1c** is the highest in energy in all three charge states. The transformation from **1a** to **2** is exothermic in all three charge states, with the energy gain being the largest in the neutral charge state and the lowest in the doubly negative charge state. From the experiment we cannot deduce the charge state in which the skeletal rearrangement from **1a** to **2** occurred.

After $N = 163$ complete dehalogenations of **1** using voltage pulses of up to 4.5 V we observed the products **3**, **4**, and **5** with occurrences of 101 (62%), 24 (15%) and 32 (20%), respectively, and in 6 cases (4%) we observed other products, shown in fig. S7. For the transitions triggered on **3**, **4**, and **5** with voltage pulses of up to +2.5 V, we did not observe other products than **3**, **4**, and **5**.

Voltage pulses of +3.5 V and higher are likely sufficient to put the molecules transiently into the anionic and dianionic charge states. Note that **3** is transiently charged doubly negative already at +1.1 V, see Fig. 2c. The intermediates **1a** and **2a** and other possible intermediates formed directly after dehalogenation could be formed in charged states and subsequent skeletal rearrangements might have proceeded in the charged molecules. The ground charge states of the products, i.e., **2** neutral, **3** anion, **4** neutral, **5** neutral, would then form after the structural transformation by charge transfer between substrate and molecule. Note that these ground states are deduced experimentally.

We calculated the energies of different fully dehalogenated products in different charge states, see Table S2, and found that the transformation from **2a** to **3**, **4** and **5** is exothermic in all three charge states. We observed in the experiment that in most of the cases after complete dehalogenation we find **3** as the initial product (62%, see Table S2), although **5** is lower in energy in neutral and negative charge state and **4** is lower in energy than **3** in the neutral charge state. However, in the doubly negative charge state, **3** is the energetically favorable structure. The dominant formation of **3** as initial dehalogenated product might therefore be explained by the intermediate **2a** being doubly negatively charged by the voltage pulse and transforming into **3** while it is in the dianionic charge state. It has been shown that the formation of sigma diradicals is thermodynamically favorable by one- or two-electron reduction [25, 37].

In the experiment structures **1a** (fig. S3A) and **2a** (fig. S7A) have been observed (very rarely). Although **2a** was found to be neutral on the surface, this does not imply that the intermediates **1a** and **2a** in the reaction pathway proposed in Fig. 1 are neutral. The reaction might have proceeded in transiently charged states. From the experiment we cannot unambiguously deduce the charge states of the short-lived intermediates.

Neutral diradicals such as **1a, 1b, 1c, 2a** are expected to be highly polarizable and possess latent zwitterionic character [38, 39]. Local electrostatic fields from Na cations and Cl anions of the environment might contribute to stabilizing these structures. Because of the large voltages needed and the irreversibility of the dehalogenation, a more detailed understanding of the reaction pathway from **1** to **3**, **4** and **5** is very challenging and out of the scope of this work.

**Details on DFT calculations**

Density functional theory (DFT) was employed using the FHI-AIMS code [40]. Each molecule, in different charge-state configurations, was investigated in gas-phase and on the surface. The geometries were optimized with the "really tight" basis defaults of FHI-AIMS. The hybrid functional B3LYP was applied [41] with van der Waals correction [42] for structural relaxation. For the different charge states, the total charge was constrained.

The convergence criterion was set at $10^{-3}$ eV/Å for the total forces and $10^{-5}$ eV for the total energies. All calculations were performed in the spin-polarized (unrestricted) framework. The structures shown in the main text in Fig. 3 O, P, Q refer to results obtained on the surface. In fig. S16 we report gas-phase energies corresponding to the different structures including transition structures for the gas phase. The transition structures where the atom positions obtained as the average between the initial and final molecular configurations.

For neutral molecules and dianions the lowest energy spin states were mostly singlets, apart from $3^0$, which was found in a triplet ground state, both on surface and in gas phase (the singlet of $3^0$ is higher in energy then the triplet by 0.83 eV in the gas phase and by 0.85 eV on the surface). The transition structures of neutral molecules and anions were singlets apart from the transition structure between $3^0$ and $5^0$, which was found in a triplet ground state (the singlet of is higher in energy by 0.58 eV in the gas phase). For all singly charged anions the ground states were doublets.

On the surface, we calculated the ground-state energies of the different structures in the different charge states. The NaCl slab was 4 layers thick. We found the same spin multiplicities as in the gas phase. On surface, $3^{-1}$ shifts further down in energy with respect to $4^{-1}$, compared to the gas phase. On surface, the molecular energies strongly depend on the adsorption site, which are not experimentally accessible for the transition structures. Calculations of the transition structures for these systems on surface are extremely demanding, requiring sophisticated methods for the localization of all possible transition structures on the different potential energy surfaces (with different charges and spin multiplicities) and therefore could not be performed in this study. For the ground state structures, i.e., $3^{-1}$, $4^0$, $5^0$, we experimentally determined the adsorption sites.

They were used as input for the on-surface calculations and the calculated structurally relaxed adsorption sites agreed with the experimentally determined ones.

Note, that we also performed calculations with the computationally less expensive PBE functional, however, this functional did not find the Jahn-Teller distorted ground state of **3$^{-1}$** (neither in gas phase, nor on surface) and therefore the PBE derived results are not further discussed in this work.

Because the charge attachment and electronic relaxation happen on a significantly faster time scale than the geometric structure relaxation of the nuclei, we also calculated for the observed electronic resonances of **3** and **4** the energies of the (transiently) charged structures in the geometry of the respective charge ground state, i.e., the initial state. In the geometry of the negative charge state (geo$^-$) of **3** we calculated the energy of the neutral **3$^0$** and the anion **3$^{-2}$**, i.e., **3$^0_{geo^-}$** and **3$^{-2}_{geo^-}$**, respectively. In the geometry of the neutral **4$^0$**, we calculated the energy of the anion **4$^{-1}$**, i.e., **4$^{-1}_{geo0}$**. The difference of such energy to that of the respective relaxed geometry corresponds to the respective relaxation energy $\Delta$ [27]. For **3$^0_{geo^-}$** we find, as for **3$^0$**, a triplet ground state (the singlet was 0.81 eV higher in energy).

**Supplementary figures**

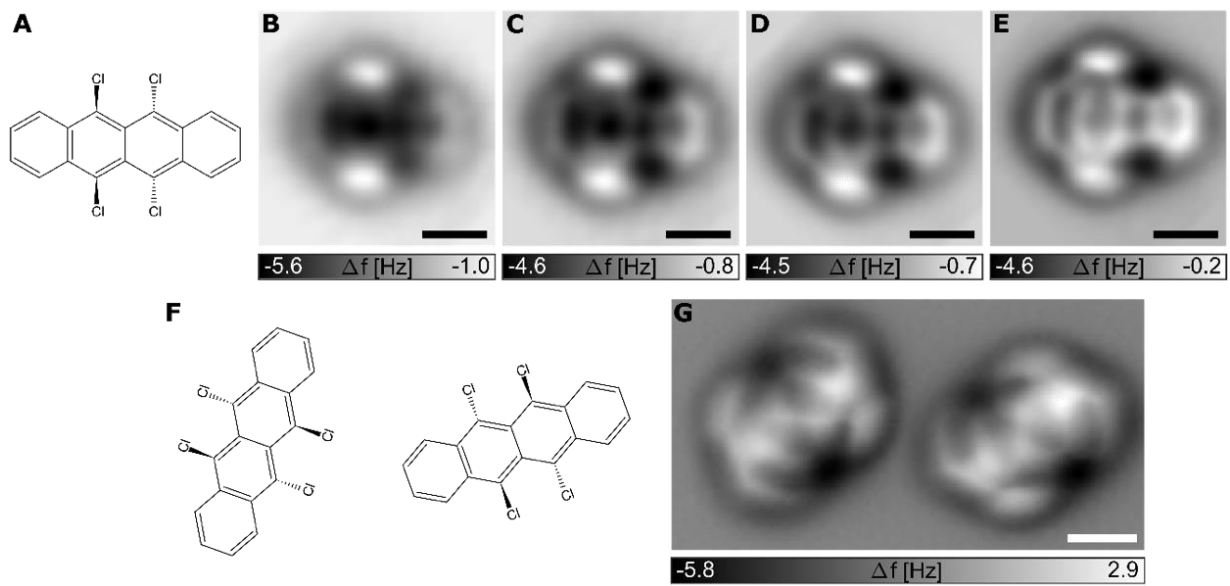

**fig. S1**: Different conformations of compound **1** recorded at different tip-height offsets: On each long side of the molecule, one Cl atom is imaged with dark contrast and one with bright contrast in AFM. The AFM contrast is related to the adsorption height [43]. We deduce that on each long side, one Cl adopts a decreased adsorption height (Cl down, imaged with dark contrast), and the other Cl adopts an increased adsorption height (Cl up, imaged with bright contrast), with respect to the tetracene backbone. This is explained by steric hindrance between neighboring Cl atoms resulting in their out-of-plane arrangement. We observe **1** in different conformations. Either the Cl atoms attached to the same carbon ring both displaced in the same direction: both Cl up on one ring and both Cl down on the other ring, i.e., *achiral*-**1**, or in different directions: one Cl up and one Cl down on each ring, i.e., *chiral*-**1**. While *achiral*-**1** is a *meso* conformation with a mirror plane of symmetry and therefore non-chiral, the latter results in a molecule with conformational chirality. (**A**) Structural drawing of *achiral*-**1**. (**B – E**) Constant-height AFM images of *achiral*-**1** recorded at tip-height offsets $\Delta z$ = 1.00 Å, 0.80 Å, 0.75 Å and 0.60 Å, respectively, with respect to a STM setpoint of $I$ = 1 pA and $V$ = 0.1 V. (**F, G**) Structural drawings of *chiral*-**1** and a corresponding AFM image recorded at $\Delta z$ = -0.25 Å with respect to the STM setpoint of $I$ = 0.5 pA and $V$ = 0.2 V. The two molecules imaged are enantiomers. All scale bars are 0.5 nm.

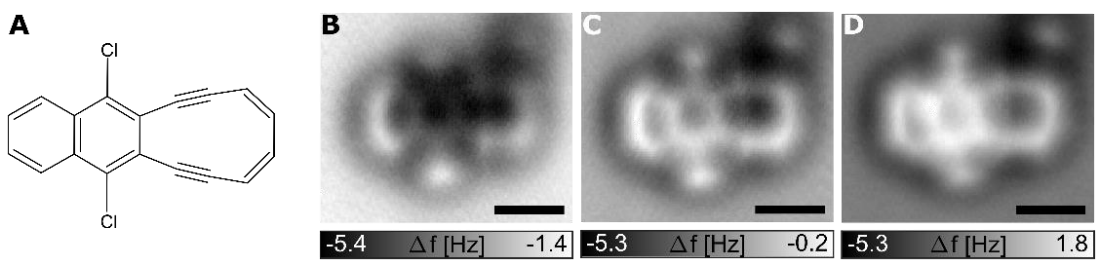

**fig. S2**: Additional AFM data on partly dehalogenated intermediate **2**. (**A**) Structural model of **2**. (**B** – **D**) Intermediate **2** imaged at tip-height offsets $\Delta z = 0.65$ Å, $\Delta z = 0.35$ Å and $\Delta z = 0.20$ Å, respectively with respect to a setpoint of $I = 1.0$ pA and $V = 0.2$ V on 2ML NaCl. Scale bars are 0.5 nm.

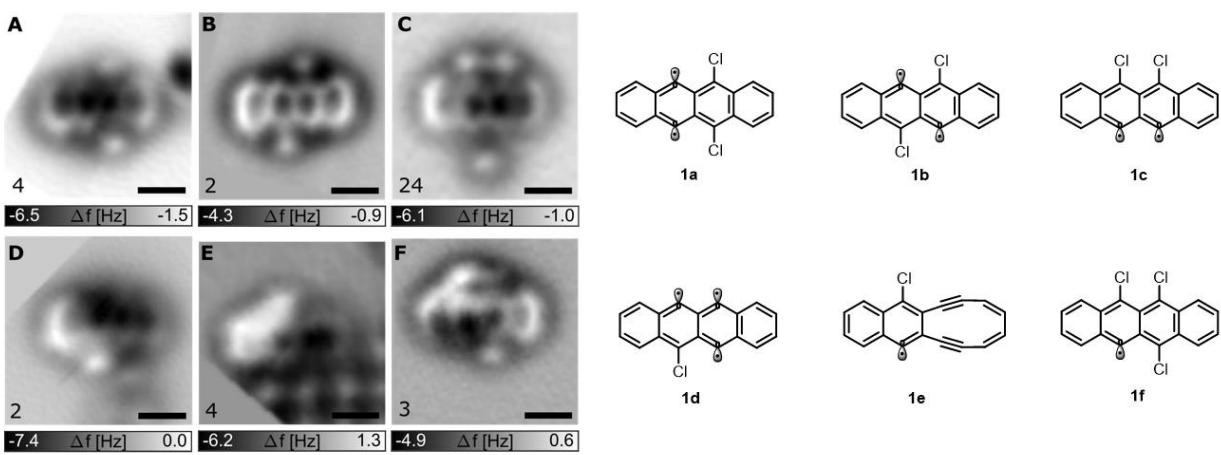

**fig. S3**. AFM images of additional, partly dehalogenated intermediates observed after applying voltage pulses of +3.5 V to +4.5 V above **1**. Occurrences of the observed species are indicated in the lower left of each image. Intermediate **2** was observed 9 times. In total $N = 48$ partly dehalogenated intermediates were observed. The scheme on the right shows the assigned respective structures. Tip-height offsets $\Delta z$ in **A-F** were 0.3 Å, 1.0 Å, 1.0 Å, 0.3 Å, 1.0 Å, and 0.5 Å, respectively, with respect to a setpoint of $I = 1.0$ pA and $V = 0.2$ V on 2ML NaCl. Scale bars are 0.5 nm.

|    | neutral charge state [eV] | negative charge state [eV] | doubly negative charge state [eV] | Occurrence in experiment in $N = 39$ intermediates observed with two Cl |
|----|---------------------------|----------------------------|-----------------------------------|------------------------------------------------------------------------|
| **1a** | 0.67 (T) | 0.27 (D) | 0.33 (S) | 4 |
| **1b** | 1.30 (T) | 0.06 (D) | **0.00** (S) | 2 |
| **1c** | 1.14 (T) | 0.64 (D) | 0.61 (S) | 24 |
| **2**  | **0.00** (S) | **0.00** (D) | 0.23 (S) | 9 |

**Table S1**: Energies of different doubly dehalogenated intermediates calculated in neutral, negative and doubly negative charge states in the gas phase with B3LYP. For each charge state, the structure with the lowest energy defines 0 eV and is used as reference for that charge state, i.e., for that column. The respective ground state's spin multiplicity is indicated in brackets with S for singlet, D for doublet or T for triplet.

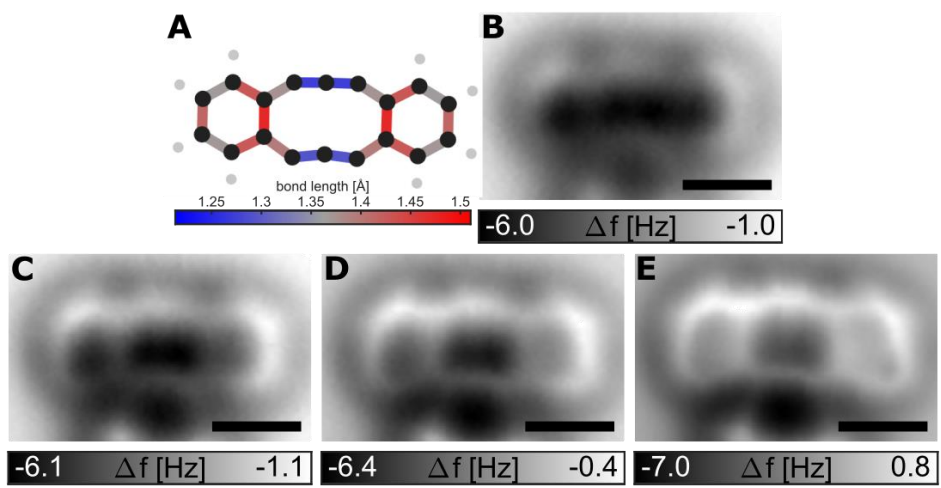

**fig. S4**. Additional AFM data on **3**. (**A**) Structural model of **3**, (**B** – **E**) Constant-height AFM images of **3** adsorbed on 2ML NaCl recorded at different tip-height offsets $\Delta z = 0.2$ Å, 0.0 Å, -0.1 Å and -0.2 Å, respectively, with respect to a setpoint of $I = 0.5$ pA and $V = 0.2$ V. Scale bars correspond to 0.5 nm.

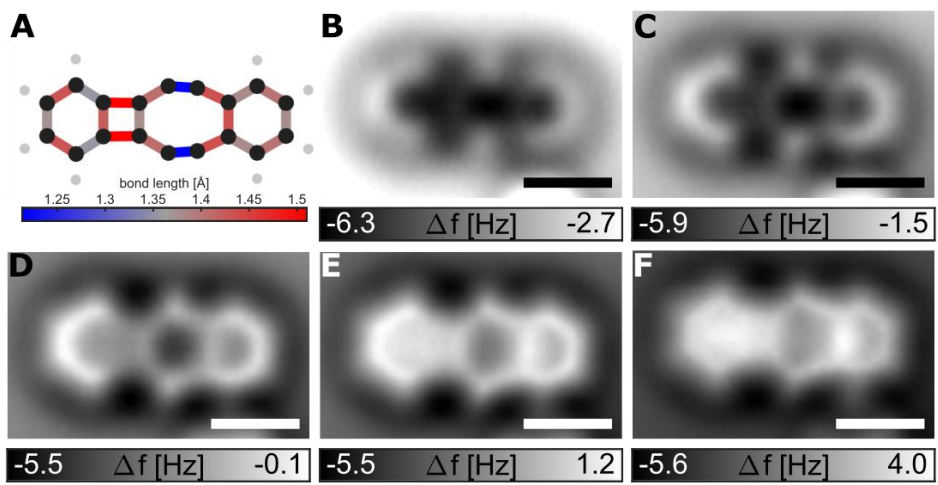

**fig. S5**. Additional AFM data on **4**. (**A**) Structural model of **4**. (**B – F**) Constant-height AFM images of **4** adsorbed on 1ML NaCl on Cu(111) recorded at different tip-height offsets $\Delta z$ = -1.2 Å, -1.4 Å, -1.6 Å, -1.7 Å and -1.8 Å with respect to the setpoint of $I$ = 1 pA and $V$ = 0.2 V. Scale bars correspond to 0.5 nm.

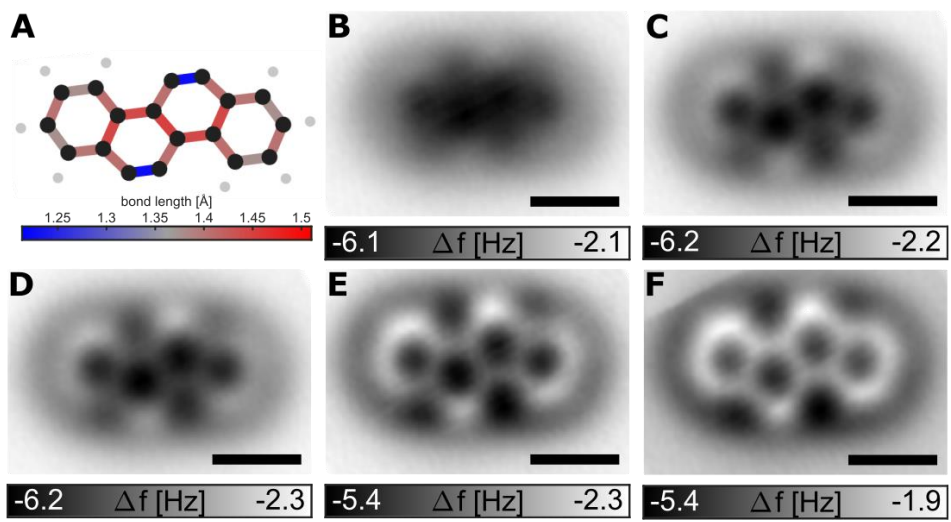

**fig. S6**. Additional AFM data on **5**. (**A**) Structural model of **5**. (**B – F**) Constant-height AFM images of **5** adsorbed on 1ML NaCl on Cu(111) recorded at different tip-height offsets $\Delta z$ = -1.3 Å, -1.5 Å, -1.6 Å, -1.7 Å and -1.8 Å with respect to the setpoint of $I$ = 0.5 pA and $V$ = 0.2 V. All scale bars correspond to 0.5 nm.

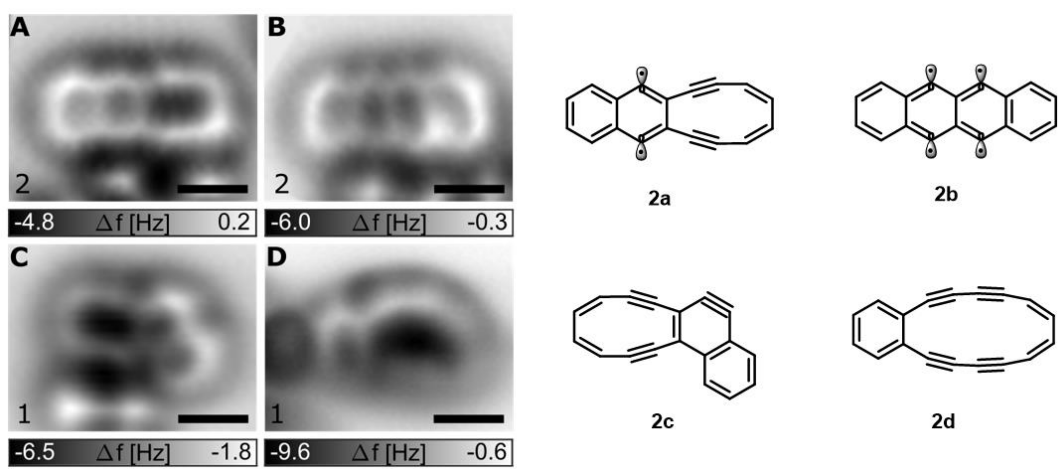

**fig. S7:** AFM images of additional products found after complete dehalogenation. The schemes to the right show the proposed respective structures. Structures **2c** and **2d** have been observed only once and are assigned only tentatively. Possibly hydrogens have been dissociated or migrated in the reactions forming these structures. Occurrences of the observed species are indicated in the lower left of each image. Structures **3**, **4** and **5** were observed as initial product after dehalogenation with occurrences of 101, 24 and 32, respectively. Tip-height offsets $\Delta z$ in **A-D** were 0.7 Å, -0.1 Å, 1.0 Å and 1.2 Å, respectively, with respect to a setpoint of $I = 1.0$ pA and $V = 0.2$ V on 2ML NaCl. Scale bars are 0.5 nm.

|  | neutral charge state [eV] | negative charge state [eV] | doubly negative charge state [eV] | Occurrence in experiment as initial dehalogenated structure in $N = 163$ events |
|---|---|---|---|---|
| **2a** | 2.60 (T) | 1.97 (D) | 1.38 (S) | 2 |
| **2b** | 3.64 (Q) | 2.73 (D) not relaxed | 2.46 (S) not relaxed | 2 |
| **2d** | 1.09 (S) | 1.17 (D) | 1.08 (S) | 1 |
| **3** | 1.19 (T) | 0.77 (D) | **0.00** (S) | 101 |
| **4** | 1.06 (S) | 0.81 (D) | 0.45 (S) | 24 |
| **5** | **0.00** (S) | **0.00** (D) | 0.29 (S) | 32 |

**Table S2**: Energies of different fully dehalogenated products calculated in neutral, negative and doubly negative charge states in the gas phase with B3LYP. For each charge state, the structure with the lowest energy defines 0 eV and is used as reference for that charge state, i.e., for that column. The respective ground state's spin multiplicity is indicated in brackets with S for singlet, D for doublet, T for triplet or Q for quintuplet. **2b** transformed into **3** in the negative and doubly negative charge state and in these charge states was calculated without structural relaxation in the geometry of **1** with all Cl atoms removed.

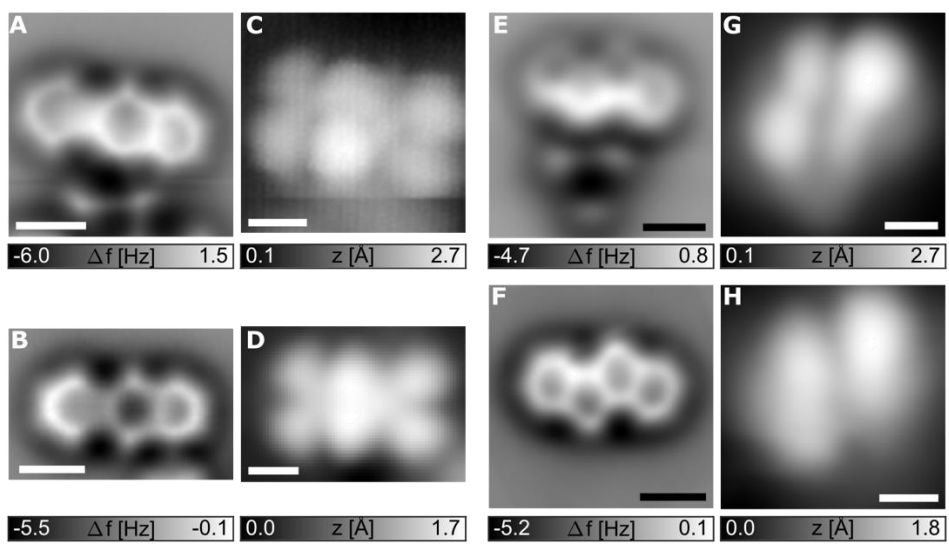

**fig. S8**. AFM and STM images of **4** (**A – D**) and **5** (**E – H**) adsorbed on 2ML NaCl (top row) and 1ML NaCl (bottom row). The contrast of the molecules shows very small differences comparing images on 2ML NaCl and on 1ML NaCl. Adsorbates (Cl adatoms, 3ML NaCl islands) near the molecules on 2ML NaCl can induce small changes in the molecular adsorption geometry; see faint asymmetry in (E) indicating an out-of-surface tilt of the molecular plane by a few degrees [43]. In (C), **4** switched to **3**, i.e., transition γ, while recording the STM image (change of contrast in the lower part of the image, slow scan direction was from top to bottom). Imaging parameters: Molecule **4**: (A) const. height AFM on 2 ML NaCl, $\Delta z = 0.0$ Å from STM setpoint $I = 0.2$ pA, $V = 0.2$ V; (B) const. height AFM on 1 ML NaCl, $\Delta z = -1.6$ Å from STM setpoint $I = 1.0$ pA, $V = 0.2$ V; (C) const. current STM on 2 ML NaCl, $I = 0.2$ pA, $V = 1.0$ V; (D) const. current STM on 1 ML NaCl, $I = 0.2$ pA, $V = 1.0$ V. Molecule **5**: (E) const. height AFM on 2 ML NaCl, $\Delta z = 0.0$ Å from STM setpoint $I = 0.5$ pA, $V = 0.2$ V; (F) const. height AFM on 1 ML NaCl, $\Delta z = -1.9$ Å from STM setpoint $I = 0.5$ pA, $V = 0.2$ V; (G) const. current STM on 2 ML NaCl, $I = 0.5$ pA, $V = 1.9$ V; (H) const. current STM on 1ML NaCl, $I = 0.5$ pA, $V = 1.7$ V. Scale bars correspond to 0.5 nm.

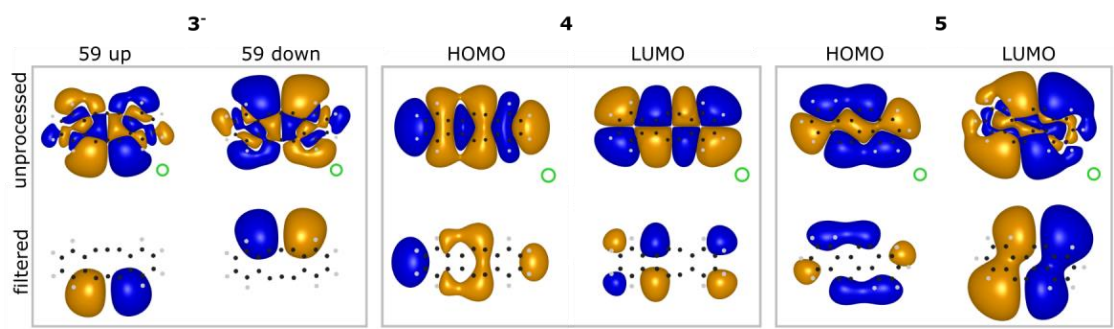

**fig. S9.** DFT derived frontier orbitals of **3**, **4** and **5**: Unprocessed DFT-derived iso-surfaces of molecular frontier orbitals of **3$^{-1}$**, **4$^0$** and **5$^0$** are shown in the top row. To simulate the finite resolution of the STM tip, these datasets were filtered using a 3D Gaussian [44]. Orbital iso-surfaces after filtering are shown in the lower row. The green ring to the bottom right of each unprocessed iso-surface indicates the full width half maximum of each 3D Gaussian filter. The nuclei positions are indicated by black and gray dots to represent carbon and hydrogen atoms, respectively. Orbitals 59 up and 59 down refer to the first occupied and unoccupied orbitals of **3$^{-1}$**, respectively, see fig. S12.

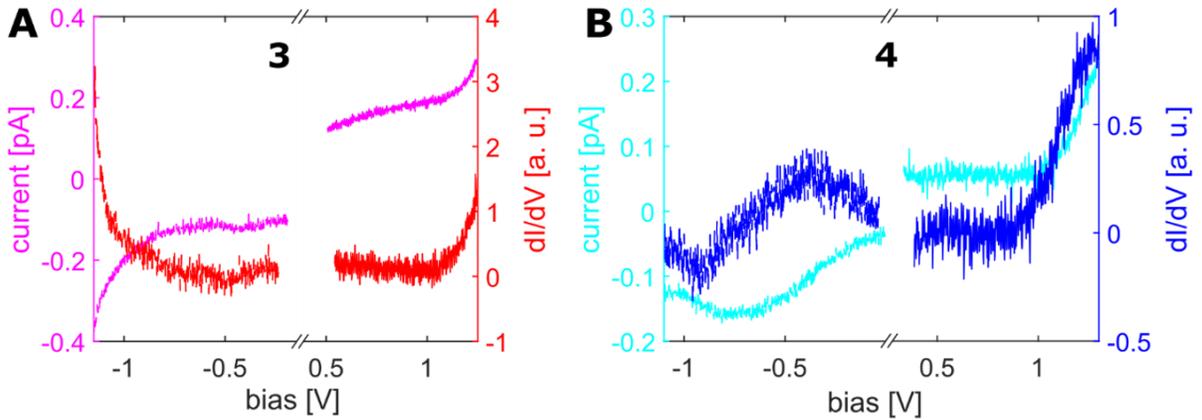

**fig. S10:** Scanning tunneling spectroscopy (STS) data on **3** (**A**) and **4** (**B**) adsorbed on 2ML NaCl on Cu(111). The onset of the electronic resonance of **3** at negative bias corresponds to the onset of structure transition α. The onset of the resonance of **4** at positive bias corresponds to the onset of structure transition γ. No resonance was observed for **4** at negative bias. Probing **4** at negative bias with constant-current STM at $I = 0.2$ pA up to $V = -2$ V, we also saw no features of an electronic resonance up to this value. Larger negative voltages could not be probed on **4** because it was then effectively converted into **3** or **5**. The faint peaks in the dI/dV spectra at $V = -0.3$ V can be assigned to the NaCl/Cu(111) interface state. The d*I*/d*V* data was obtained by numeric differentiation of the *I*/*V* data.

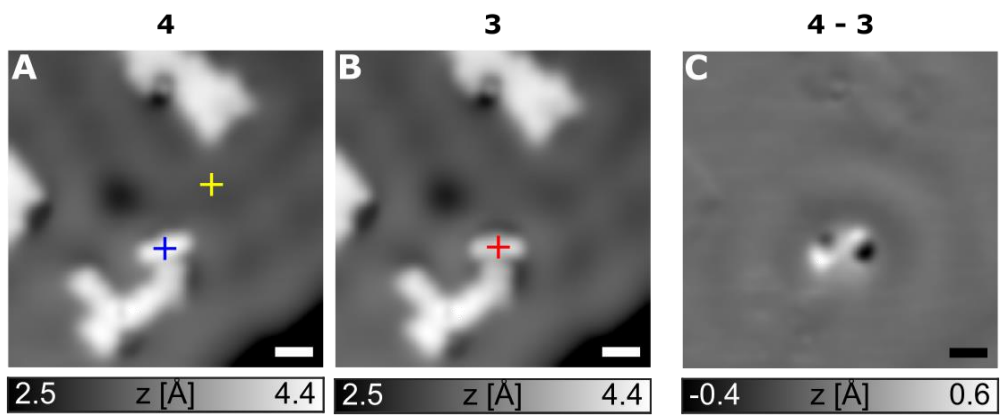

**fig. S11**. Interface-state scattering of **3** and **4** on 2ML NaCl. (**A**) and (**B**): Constant-current STM images at $V = 50$ mV, $I = 0.5$ pA. (**C**) shows the difference image, that is, (A) minus (B). The yellow, blue and red cross indicate the tip position while recording Kelvin probe force spectroscopy (KPFS) data on **4**, **3** and on bare bilayer NaCl, respectively. All scale bars correspond to 1 nm. The observed scattering patterns around the molecule in the difference image indicate that the charge state of the molecule is changed with the structural transition between **3** and **4** [31, 45]. The more pronounced scattering pattern around **3**, compare (A) and (B) suggests that **3** is the charged species, which is corroborated by the KPFS measurements [46, 47], shown in Table S3.

| $\Delta z$ [Å] | LCPD above **3** [eV] | LCPD above **4** [eV] | LCPD above 2ML NaCl [eV] |
|---|---|---|---|
| 3.0 | 0.366 | 0.308 | 0.256 |
| 2.0 | 0.375 | 0.311 | 0.237 |
| 1.0 | 0.427 | 0.322 | 0.214 |
| 0.5 | 0.437 | 0.361 | 0.205 |

**Table S3**. Kelvin probe force spectroscopy (KPFS) on **3** and **4** on 2ML NaCl. The values show the Local contact potential difference (LCPD) in eV extracted form KPFS data by fitting the peak position of a parabola, recorded with open z-feedback at tip-height offset $\Delta z$. Tip positions are indicated in Fig. S11. The $\Delta z$-values refer to the STM setpoint of $I = 0.5$ pA and $V = 200$ mV on 3ML NaCl. In the entire $\Delta z$-range, the LCPD above **3** is larger than above **4**, indicating that **3** is more negatively charged then **4**.

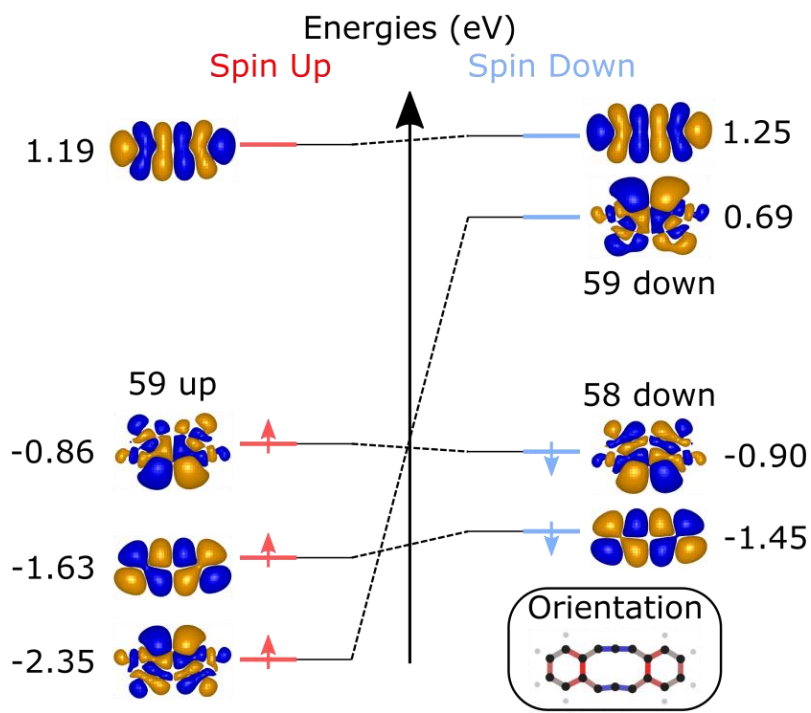

**fig. S12.** DFT calculated frontier orbitals and energies of the anion of **3** in the gas phase using B3LYP. The anion of **3** is a radical and features an S = ½ ground state. The Jahn-Teller distortion of **3⁻¹**, i.e., an inside bend (see inset), breaks the symmetry of the molecule, which also is reflected in the orbital densities. The highest occupied states, i.e., the first occupied state, 59 up, and the almost energy degenerate second occupied state, 58 down, both show two lobes with increased density above the side with the bend. The first unoccupied state, 59 down, corresponds to the spin minority (down) and shows two lobes with increased density above the side opposite to the bend. The bend is resolved with AFM (see Fig. 3M and fig. S3) and the position of the dominant lobes at occupied and unoccupied states agree with STM images at the electronic resonances (see Fig. 3B, C).

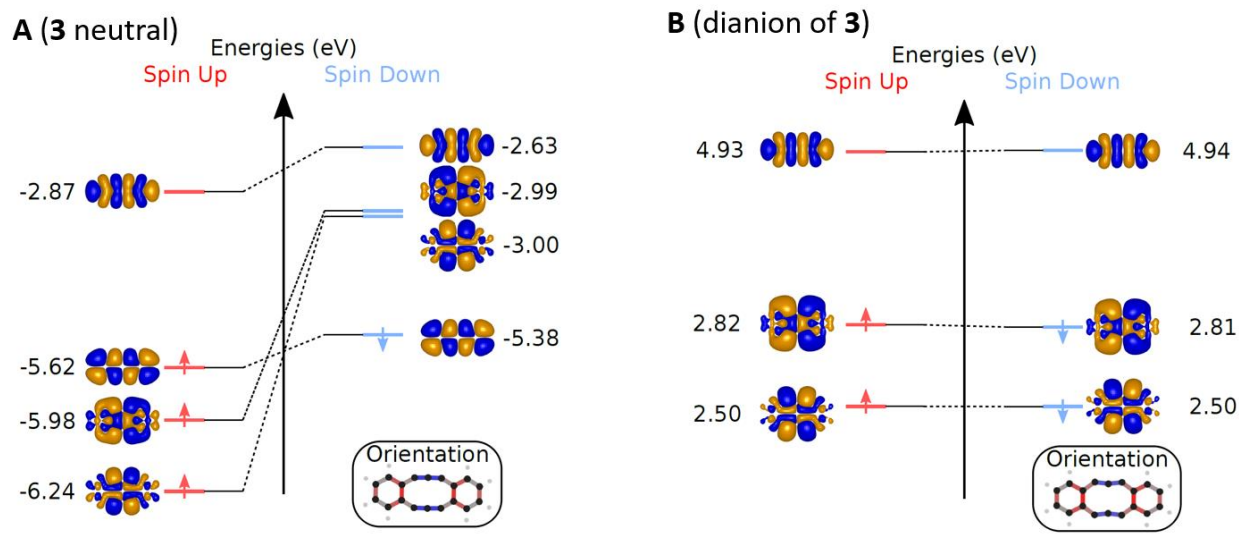

**fig. S13.** DFT calculated frontier orbitals and energy spectra of **3** in neutral (**A**) and in the doubly negative charge state (**B**) in the gas phase using B3LYP. In the neutral form **3** is found in a triplet ground state. Note that in neutral and doubly negative charge state the relaxed molecular structure and the orbital densities are mirror symmetric with respect to the vertical plane through the long axis of the molecule, in contrast to the Jahn-Teller distorted anion (see fig. S12). The orbital densities for neutral and doubly negative charge states differ significantly from the experimentally observed ones (see Fig. 3B, C), corroborating the assignment of the negative charge ground state of **3** on the surface, presented in fig. S12 and Table S3.

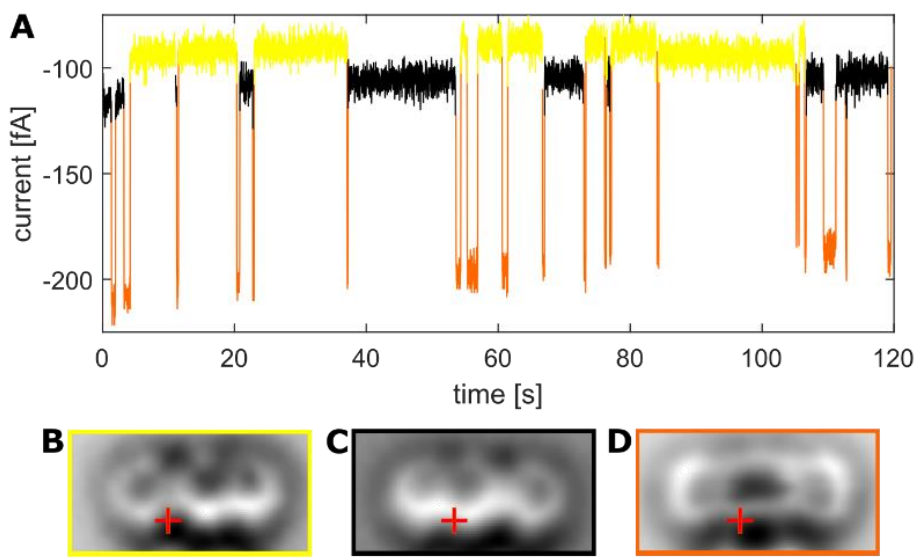

**fig. S14.** (**A**) Current trace $I(t)$ with switching events between **3** and **4** at constant tip height at $V = -1.3$ V. Three current plateaus are visible, corresponding to different structures. Multiple structural transitions occurred during a time of 120 s. Typically molecules moved in conjunction with induced structural transitions, but in the experiment shown here a molecule was stably anchored at a 3$^{rd}$ layer NaCl island edge, which is at the bottom side of the AFM images (**B**, **C**, **D**). The three current plateaus are assigned by reducing the voltage at a certain plateau and recording an AFM image of the corresponding structure at $V = 0$ V, where no switching occurs, and structures are stable. Such AFM images, corresponding to the colors of the respective plateaus are shown in (B, C, D). The red crosses indicate the tip position at which the current trace was recorded. Multiple switches from **3** to **4** (transition α) and from **4** to **3** (transition β) occurred. In the current trace structure **3** (D) and the two mirror symmetric structures of **4** are discerned (B, C). For the statistics shown in Fig. 4, we did not differentiate between the two mirror symmetric structures of **4**. For structure **3**, shown in (D), we did not observe the corresponding mirrored structure, which would have the inward bend related to the Jahn-Teller distortion on upper side of the molecule. Observing only one of the distorted possible structures is expected as there is always some anisotropy due to the environment in a real system, which will determine the orientation of the Jahn-Teller distortion, as shown for other systems [48, 49]. In general, we observed the bend of **3** on the side next to a 3$^{rd}$ layer NaCl island, see panel (D). Note, that the position of the bend is not only inferred from AFM data, but independently also from the orbital resonances of **3** imaged by STM (see Fig. 3B, C and fig. S12).
31

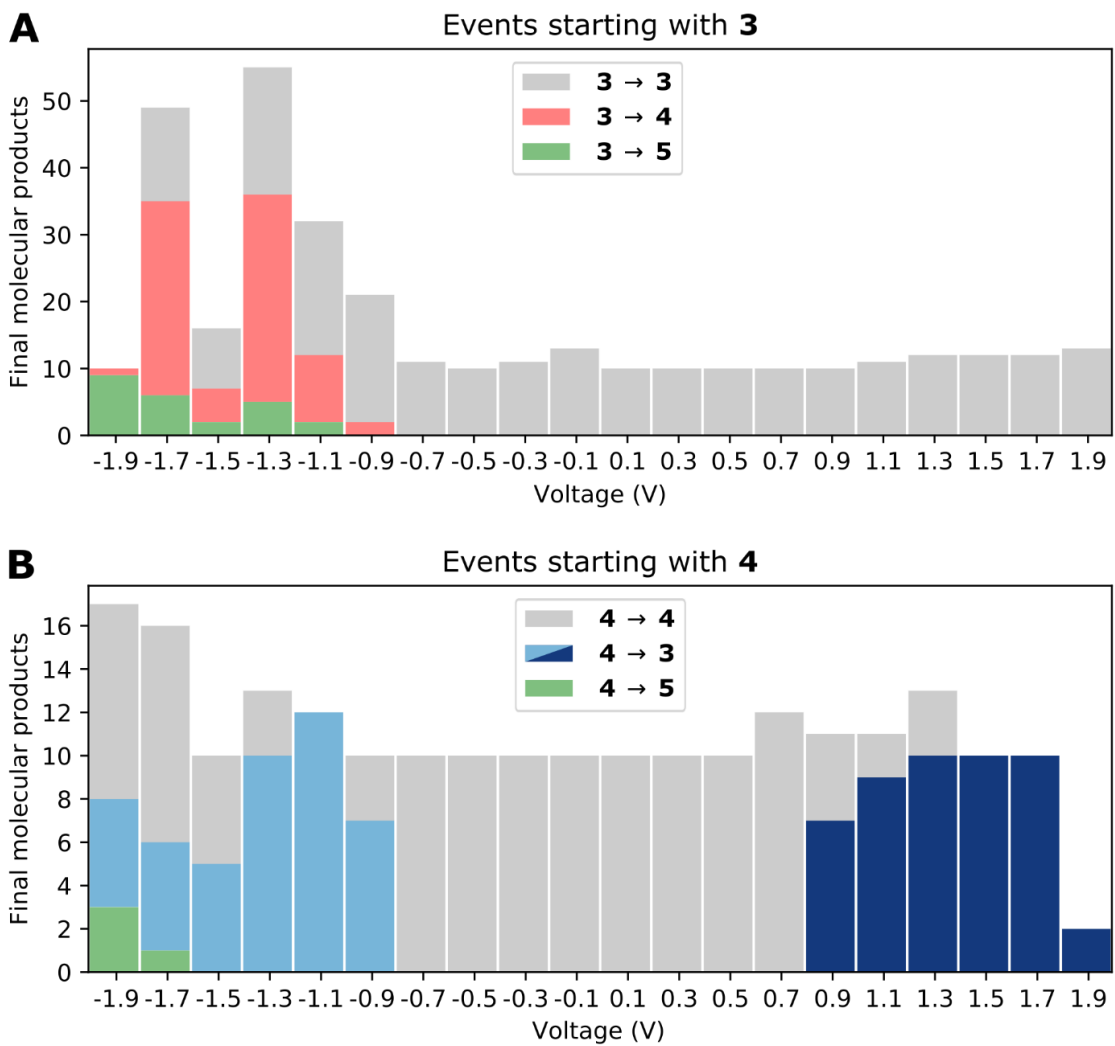

**fig. S15:** Total number of events and trials for applying pulses with different voltages *V* in the range of *V* = -1.9 V to 1.9 V at a constant current of *I* = 0.5 pA for *t* = 10 seconds starting with **3**, see (**A**), and starting with **4**, see (**B**), shown here and in Fig. 4A, B. In total, we analyzed *N* = 555 voltage pulses on 44 individual molecules for the investigation of the voltage dependence shown here and in Fig. 4A, B.

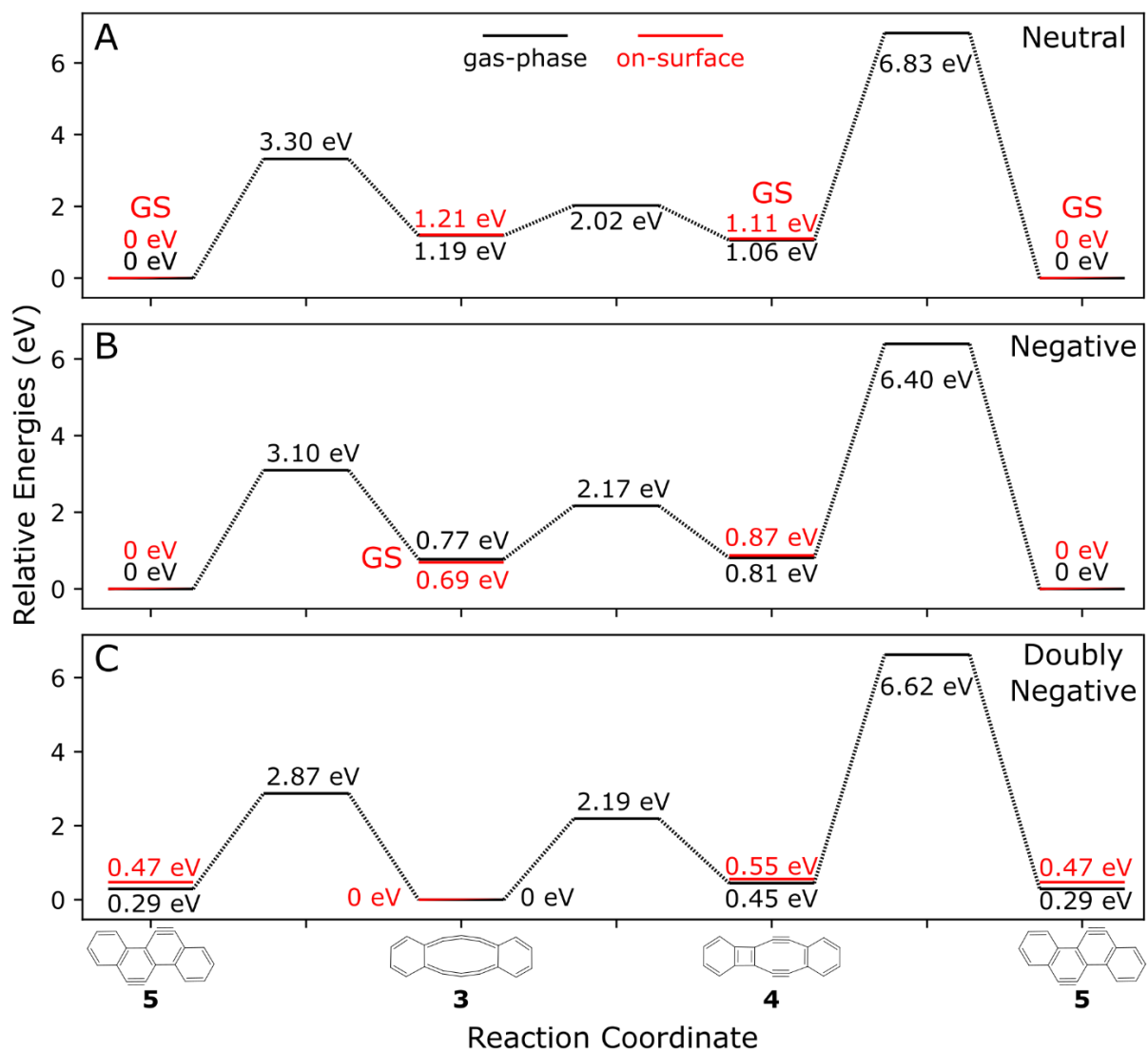

**fig. S16.** Energies of structures of **3**, **4** and **5** in different charge states calculated in the gas phase (black, including transition structures) and on NaCl (red, without transition structures). (**A**): neutral, (**B**): negative, (**C**): doubly negative. The experimentally determined charge ground-state of each structure, i.e., neutral for **5** and **4**, but singly negative for **3**, are indicated by "GS". Energies of transition structures were not calculated on surface, because the adsorption geometry and adsorption site of the transition structures is not known.

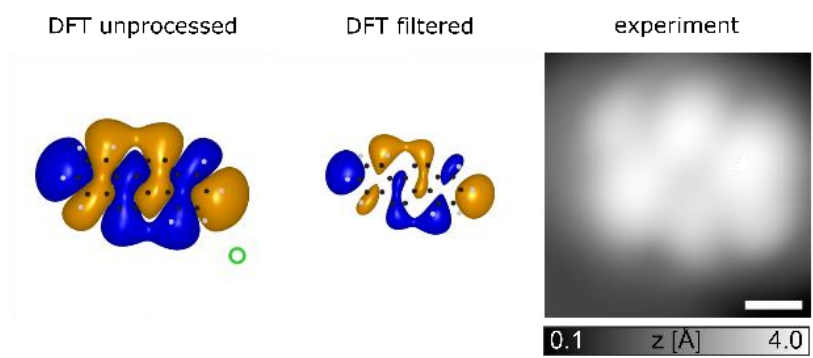

**fig. S17:** DFT-derived LUMO+1 of **5** without (left) and with (middle) applying a low pass filter to mimic the finite lateral resolution of the STM tip. The experimental orbital density map (right) was recorded with const. current STM at $I = 0.5$ pA and $V = 2.5$ V. The scale bar corresponds to 0.5 nm. Switching from **5** to **3** or **4**, typically required currents of $I > 7$ pA at $V = 2.5$ V.